# Real-Time Monitoring of Undirected Networks: Articulation Points, Bridges, and Connected and Biconnected Components*

Giorgio Ausiello    Donatella Firmani    Luigi Laura

Dipartimento di Informatica e Sistemistica, Sapienza University of Rome.
Via Ariosto, 25. 00185 Rome, Italy. {ausiello,firmani,laura}@dis.uniroma1.it

November 17, 2016

### Abstract

In this paper we present the first algorithm in the streaming model to characterize completely the biconnectivity properties of undirected networks: articulation points, bridges, and connected and biconnected components. The motivation of our work was the development of a real-time algorithm to monitor the connectivity of the Autonomous Systems (AS) Network, but the solution provided is general enough to be applied to any network.

The network structure is represented by a graph, and the algorithm is analyzed in the datastream framework. Here, as in the *on-line* model, the input graph is revealed one item (i.e., graph edge) after the other, in an on-line fashion; but, if compared to traditional on-line computation, there are stricter requirements for both memory occupation and per item processing time. Our algorithm works by properly updating a forest over the graph nodes. All the graph (bi)connectivity properties are stored in this forest. We prove the correctness of the algorithm, together with its space ($O(n \log n)$, with $n$ being the number of nodes in the graph) and time bounds.

We also present the results of a brief experimental evaluation against real-world graphs, including many samples of the AS network, ranging

---

from medium to massive size. These preliminary experimental results confirm the effectiveness of our approach.

# 1 Introduction

Studying the connectivity properties of an undirected graph is the first step in its basic structural analysis. A *connected component* of a graph is a (maximal) set of nodes that can be mutually reached, and a graph is said to be connected if it consists of a single connected component. The *articulation points*, also known as *cut-vertices*, and the *bridges* are, respectively, nodes and edges whose removal disconnects the graph, if it is connected, or increase the number of connected components otherwise. A (maximal) set of nodes is a *biconnected component* if there are at least two distinct paths connecting each pair of nodes.

Computing the above properties in the traditional off-line setting is a problem that dates back to the seventies, when the first algorithms appeared in the classical works of Hopcroft and Tarjan [16] and Tarjan [25], that run in linear time, and are based on depth-first search. If we switch to the on-line setting, the first algorithms to maintain the bridge-connected and the biconnected components have been addressed by Westbrook and Tarjan [27] already in 1989. The authors developed an efficient data structure for this problem based on disjoint-set data structures. Specifically, their algorithm processes $n$ node additions and $m$ edge additions in $O(n + m\alpha(m, n))$ total time and $O(n \log n)$ space, where $\alpha$ is the functional inverse of Ackermann's function, a very slowly growing function[1].

This time bound is proved to be optimal. As we will see, our approach is not far from the classical approach but with a few remarkable aspects that we describe in detail in Section 6.

Our work was inspired by the problem to find *bridges* and *articulation points* in the *datastream* framework to analyze the topology of the net of *Autonomous Systems* and to discover what nodes and links are "critical". So in our scenario the first requirement was to make a query on a link and respond in $O(1)$ time, as we will explain in detail in Sections 6 and 7. We wanted also a single algorithmic solution to track both properties during the monitoring of the net.

---

[1] For every "practical" value of $m$ and $n$, it holds that $\alpha(m, n) \leq 4$.

Within the Internet, an Autonomous System (AS) is a collection of connected Internet Protocol (IP) routing prefixes, under the control of one or more network operators; an Autonomous System presents a common, clearly defined routing policy to the Internet, and uses BGP (Border Gateway Protocol) to connect to other ASes. Each AS is assigned a unique Autonomous System Number, for use in BGP routing. The AS number is important because it uniquely identifies each network on the Internet. Announcements BGP (available connecting to publicly accessible looking glass servers) generate a datastream of routing paths that can be simply parsed into a datastream of links between ASes. In the literature ASes have been studied extensively, from several points of view, including their network structure [13], their mutual relationship [14], their degree distribution [10], and routing algorithms (see, e.g., [6]).

We propose a *sketch* data structure and an algorithm, called *At First Look* (AFL), that computes *all* the properties mentioned above of a streaming graph. The key idea is to keep in memory a particular spanning forest of the input graph - a forest that we call a *navigational sketch* and will define in Section 3 - in which maximal trees represent connected components, edges are distinguished in two types (solid, representing bridges, and colored, representative of the biconnected components), and the articulation points are distinguished by the types of incident edges.

We designed this algorithm in the *datastream* framework: here, as in the on-line framework, the items (graph edges) arrive one after the other, but there are stricter requirements concerning memory occupation and the allowed processing time of a single item/edge (PIPT: Per Item Processing Time), that represents the capacity of a datastream algorithm to deal with a high rate stream of data.

As we will discuss in Section 7, the amortized PIPT of our algorithm is constant if the graph is slightly more dense than a regular graph, i.e., the edge-node ratio is greater than or equal to $O(\log n)$, and the results of the experimental evaluation on both worst case and amortized PIPT, reported in Section 8, confirm the effectiveness of our approach.

Furthermore, we provide an easy to implement data structure, and we prove that the overall processing time, on a graph with $n$ nodes and $m$ edges, is $O(n \log n + m\alpha(m, n))$, which is very close to the lower bound provided in [27]. The space usage of that data structure is $O(n \log n)$, where $n$ is the number of nodes of the graph, which is the space requirement for the so-called *semi-streaming* graph algorithms; note that the space bound is tight, in the sense that, for particular graph instances, it is the space needed to store part of the solution (the list of all the bridges).

Other sections of this paper are organized as follows: in the next section we discuss related works, while preliminary issues are addressed in Section 3; in Section 4 we present the proposed algorithm, whose correctness is proved in Section 5. The data structure and its analysis are detailed, respectively, in Sections 6 and 7. The experimental results are described in Section 8, whilst concluding remarks are discussed in Section 9.

## 2 Related Work

In *classical streaming*, implicitly defined in the early work of Munro and Paterson [19] and later diffusely adopted (see e.g. [15, 20]), the input is a data stream, to be accessed sequentially (in an adversarial order), and to be processed with a working memory that is small with respect to the length of the stream. The key parameters of this model are the number of *passes $p$* and the memory *size $s$*, together with the *per item processing time* that must be kept small if there is a real-time constraint.

The restrictions imposed by classical streaming proved to be too strict to allow efficient solution for basic graph problems such as *connectivity* and *shortest paths* [15], and Feigenbaum et al. [12], exploiting the idea originally introduced by Muthukrishnan [20], proposed the *semi-streaming* model, in which the working memory size is $O(n \text{ polylog } (n))$, where $n$ is the number of nodes of the streaming graph: as in the semi-external memory model [1, 26], the main memory allows one to store data related to the nodes but not to the edges; Muthukrishnan [20] defines this memory requirement as a "sweet spot" for graph problems, and in this model several results appeared recently, including: *connected components*, *bipartiteness*, *bipartite matching*, *minimum spanning tree* [12, 11], *triangle counting* [3], *matching* [17], and *t-spanners* [2, 9, 11]. In particular, in the work of Feigenbaum et al. [12], the authors present also an algorithm to compute the articulation points of a graph, but, since it uses a disjoint set data structure for each node of the input graph (to store the node's neighbors), its memory space requirements are not apparently within the bounds of the semi-streaming model.

The interested reader can refer, as starting points on general problems in streaming models, to the book of Muthukrishnan [20] and to the recent survey of Demetrescu and Finocchi [8]. For graph problems, useful references are the Ph.D. Thesis of Ribichini [22], and the entry in the Encyclopedia of Database Systems written by McGregor [18].

## 3  Statement of the Problem

In this section we recall a few definitions from graph theory, introduce some others that we will use in the explanation of the proposed algorithm and present the statement of the problem.

Given a graph $G = (V, E)$, we define:

- **connected component**: a maximal set of nodes $V' \subseteq V$ such that, given $u, v \in V'$, there is at least one path between $u$ and $v$ in $G$;

- **articulation point**: a node $v \in V$ such that its removal from the graph $G$ increases the number of connected components in $G$;

- **bridge**: an edge $e \in E$ such that its removal from the graph $G$ increases the number of connected components in $G$;

- **biconnected component**: a maximal set of nodes $V'' \subseteq V$ such that, given $u, v \in V''$, there are at least two distinct paths between $u$ and $v$ in $G$;

In Figure 1 we can see an example of a graph in which all the above defined elements are shown. It is important to note that all the nodes adjacent to bridges are articulation points unless the bridge is their only incident edge, as is the case for node $A$ in Figure 1: the removal of the bridge leaves $A$ isolated, but the removal of $A$ (together with all its adjacent edges, i.e., only the bridge) does not increase the number of connected components of the resulting graph.

Let us now define the object that we call a *navigational sketch NS* of a graph $G$ as follows:

**Definition 1** *Given a graph $G = (V, E)$, its navigational sketch is a graph (forest) $NS$, where the set of nodes is the same of $G$, i.e., $V_{ns} = V$, and the set of edges $E_{ns}$ is distinguished in two types:* solid *and* coloured *edges. The following properties hold:*

1. *the* connected components *of $G$ are the maximal trees in $NS$;*

2. *the* bridges *of $G$ are the* solid *edges of $NS$;*

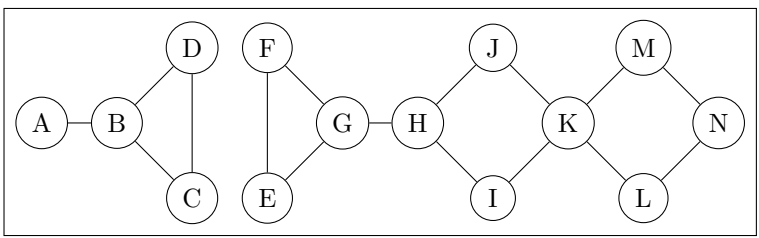

Figure 1: An example graph with two connected components (nodes A..D and E..N), four articulation points (B,G,H,K), two bridges (edges (A,B) and (G,H)), and four biconnected components ((B,C,D), (E,F,G), (H,I,J,K), and (K,L,M,N)).

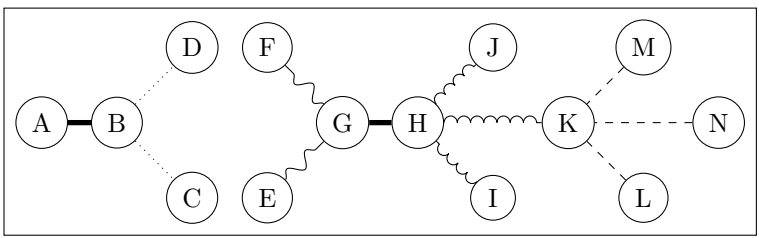

Figure 2: A navigational sketch of the graph in Figure 1: with the line style ▬▬ we represent solid edges, while colored ones are represented, respectively, by the styles ........, ∿∿, ⌒⌒, and - - - -.

3. the biconnected components *of G are represented with a subtree, inside a tree in NS, with one father and $b-1$ children (where b is the cardinality of the biconnected component); all the edges in the subtree are of the same color, and this color is unique inside NS.*

In Figure 2 we can see an example of a navigational sketch of the graph shown in Figure 1. It is important to note that, given a graph $G$, the navigational sketch of $G$ is not unique; furthermore, given a navigational sketch $NS$, there are several graphs such that their navigational sketch is $NS$: the navigational sketch of a graph of $G$ is, indeed, a real *sketch* of $G$; we can use a navigational sketch to compute the (bi)connectivity properties of a graph but other properties, such as the distance between two nodes, are not maintained by the navigational sketch.

As we can see from the above definition, many of the (bi)connectivity properties of $G$ are mapped into the navigational sketch; we miss the articulation points but, before detailing how to compute them, it is important to emphasize the role of the two distinct edge types in the NS:

1. **SE**: solid edges, that are real edges of $G$ and correspond to its bridges;

2. **CE**: coloured edges, that are representative of *biconnected components* of $G$ (i.e., all the nodes connected by edges of the same color belong to the same biconnected component).

We define the *colour degree* of a node as follows:

**Definition 2** *Given a graph $G = (V, E)$, its navigational sketch $NS$, and a node $i \in V$, the* colour degree *of $i$, denoted by $d_c(i)$, is equal to the number of incident solid edges plus the number of distinct colors of incident colored edges.*

For example:

- a node $i$ adjacent to ten edges of the same color has $d_c(i) = 1$;

- a node $i$ adjacent to two solid edges, three red edges, and two yellow edges, has $d_c(i) = 4$, i.e., two solid edges, plus two distinct colors (red and yellow).

From the above definitions we can now characterize also the articulation points of $G$, as stated in the following lemma:

**Lemma 1** *Given a graph $G = (V, E)$ and its navigational sketch $NS$, each node $u \in V$, such that $d_c(u) > 1$, is an articulation point of $G$.*

**Proof** If $d_c(u) > 1$, it is easy to see that one or more of the following properties of $u$ hold true:

1. $u$ belongs to more than one biconnected component;

2. $u$ belongs to at least a biconnected component and it is adjacent to (at least) a bridge;

3. $u$ is adjacent to more than one bridge;

Each of the previous properties define $u$ as an articulation point. □

We define the problem studied in our work as follows:

**Problem 1** *Given a streaming graph $G$ represented by a stream of its edges $S = e_1, e_2, \ldots, e_m$ (in any order), the goal is to compute all its (bi)connectivity properties: connected components, articulation points, bridges, and biconnected components.*

In order to solve this problem, as we will see in the following sections, the idea is to maintain a navigational sketch of the streaming graph; in particular, after each item (edge) from the stream is processed, we properly update the navigational sketch. We prove the correctness of the this approach in Section 5.

## 4 "At First Look" Algorithm

We now provide a high level view of the algorithm, while missing details will be covered later. As already mentioned in Section 1, the main idea behind the algorithm "At First Look" (AFL) is to keep in main memory a *navigational sketch* of the input graph $G$. Let us now take a closer look to understand how it can be built and used by the algorithm.

The set of operations to build and maintain the navigational sketch are:

**op-A.** find whether two nodes are in the same tree;

**op-B.** join trees;

**op-C.** find whether nodes are extremes of same coloured edges;

**op-D.** join sets of same coloured edges or solid edges;

**op-E.** find the path joining nodes.

A data structure that represents a navigational sketch and supports this set of operation consistently with the time and space boundaries imposed by the datastreaming model (ref. Section 1), will be discussed and analyzed in Section 6. In the present section, to provide a high level view of the algorithm we simply assume the existence of the structure and the operations that it supports.

As we can see in the pseudocode of AFL, shown in Algorithm 1, at each step the algorithm looks at the current edge from the stream and, *at first look*, it decides on the corresponding action:

1. if the current edge joins two different trees (i.e., connected components), it is added to the forest as a *solid edge*: indeed, it is the only (so far) edge to connect two distinct connected components, and therefore is a bridge;

2. if the edge connects two nodes inside the same connected component, we need to distinguish two cases (we recall that, inside a tree, there is a unique path joining two nodes) according to the path between the two nodes:

    (a) all the edges in the path are of the same color: this implies that they already are in the same biconnected component, and therefore we can drop this edge (i.e., no change in the forest);

    (b) the path is made of different edge types (e.g., some solid edges and some colored edges, or colored edges with at least two different colors): this means that, together with the current edge, all the *involved* nodes form a biconnected component and the tree needs to be changed accordingly.

---

**Algorithm 1** Algorithm *At First Look.*

---

For each edge *(i,j)* of the stream:

- test whether $i$ and $j$ are in the same tree (**op-A**)

    1. **$i$ and $j$ are in different trees:** join the trees with a *solid edge* (**op-B**);
    2. **$i$ and $j$ are in the same tree:**
        - find the (unique) path in the navigational sketch connecting $i$ and $j$ (**op-E**);
        - test whether all the edges in this path have the same colour (**op-C**):
        
        **2(a). all the edges in this path have the same colour:** ignore *(i,j)*;
        
        **2(b). the edges in this path have distinct colors and/or some/all of them are solid:** join the set of same coloured edges and solid edges "touched" by the path using the Algorithm 2 (**op-D**).

---

Let us clarify the last action, i.e., case 2 (b), that is probably the most delicate step: if we are in this case that means that, with the current edge, we found a second path, inside the tree, connecting a set of nodes; therefore we

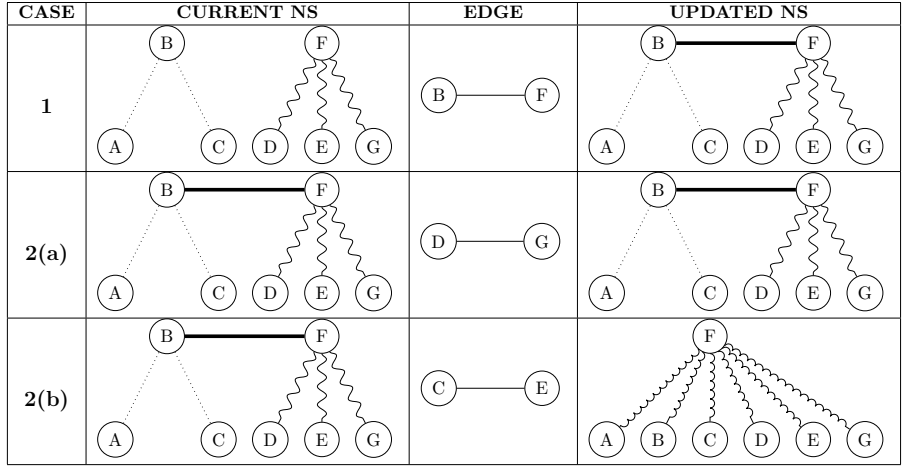

Figure 3: Example of the three cases of Algorithm AFL: for each case we show the (current) structure of the navigational sketch (i.e., before the processing of the current edge from the stream), the current edge (from the stream), and the updated structure of the navigational sketch (after the processing of the current edge).

**Algorithm 2** Algorithm to join edges in a path.

---

Given the unique path $P$ (in the navigational sketch) that connects node $i$ to node $j$, consider the set consisting of the following edges:

- solid edges that belong to $P$;
- coloured edges that belong to $P$;
- coloured edges not belonging to $P$ and coloured with the same colour of any of the coloured edges belonging to $P$.

Execute the following operations:

- delete each edge in one of these three categories;
- let be $n$ the number of nodes involved in the deletion process, i.e., every node adjacent to one of the edges described before. Now add $n-1$ new coloured edges with a new colour, representing the new biconnected component just found, such that the resulting tree structure has one father and $n-1$ children.

---

need to update the tree structure in order to reflect the fact that we found a new biconnected component: all the biconnected components "touched" by the path will be merged into a new, single biconnected component. In order to do so, as shown in Algorithm 2, we i) consider the set of all the involved nodes, that are the nodes in the path together with all the nodes connected by edges of the same colors of at least one edge in the path; ii) remove all the edges between the involved nodes, iii) choose a node, amongst the involved nodes, to be the "father", and iv) insert an edge between the previously chosen node and all the other involved nodes.

In Figure 3 we can see a graphical example of the three cases of the AFL algorithm; as we can see, for the case 1, we add a solid edge in the navigational sketch, in case 2 (a) the navigational sketch remains unchanged, whilst in case 2 (b) the structure of the navigational sketch changes significantly. In the Appendix an execution of the AFL algorithm is shown against a simple graph: we detail every step with a snapshot of the navigational sketch.

Note that the above described procedure does not alter the tree structure: it takes a tree and, after its execution, the resulting graph is still a tree.

## 4.1  Comparison with previous work

As we mentioned in the introduction, it is worth comparing our approach to the one proposed by Westbrook and Tarjan [27]. In particular, they study two distinct problems closely related to the one considered in this paper:

1. the maintenance of *bridge-blocks*, or bridge connected components, that are the components of the graphs formed by deleting all the bridges;

2. the maintenance of *blocks*: a block is a biconnected component or, if an edge is not contained in any cycle of the graph, then the single edge is a block (note that, if this is the case, it is also a bridge).

In Figure 4 we see a graph $G$, together with its bridge-blocks and blocks. In order to maintain the bridge-blocks and blocks, the authors use two distinct tree structures:

1. *bridge-block forest* (BBF) is a collection of bridge-block trees (BBF), and the nodes of a BBF are of two types: square nodes, which represents the nodes of $G$, and round nodes, which represents the bridge-blocks. Every square node is a leaf, and it is connected to the square node that represents the bridge-block it belongs to. Square nodes are connected by bridges.

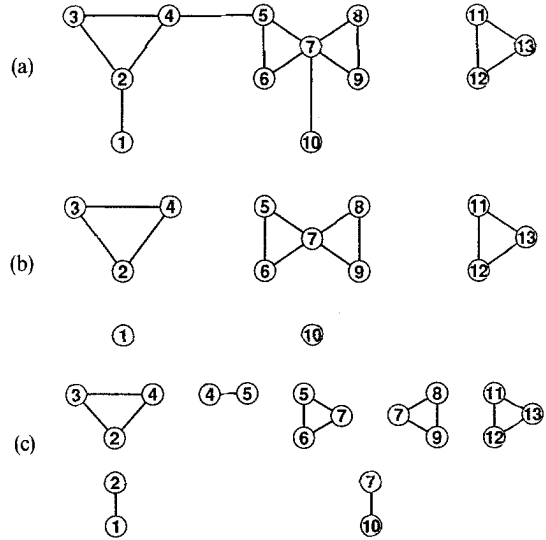

Figure 4: An undirected graph $G$ (a). Bridge-blocks of $G$ (b). Blocks of $G$ (c). From [27].

2. *block forest* (BF) is a collection of block trees (BT); each block is a round node and each node of the graph is a square node. Different from BBF, the graph (tree) is bipartite: round nodes can be connected only to square nodes.

Although the authors did not consider explicitly either *articulation points* or *bridges*, it is simple to see that the BBF tree can easily be modified in order to maintain information about the bridges: since every edge connecting two round nodes is a bridge, it suffices to store the information about which nodes are adjacent to this edge.

Furthermore, also the BF tree can be modified to maintain information about both bridges and articulation points: each round node connected to only two square nodes is a bridge (between those nodes), and each square node connected to (almost) two round nodes is an articulation point.

Both the above structures are implemented using a sophisticated data structure, called a *link/condense tree*, derived from the dynamic tree data structure of Sleator and Tarjan [23, 24], and capable of supporting fast path-finding and node and path condensation. The authors provide two significantly different versions of the *link/condense tree*: one, for the BBF, that supports condensation for every node in a path, and the other, for the BF, that supports condensation for every other node in a path, since only alternating nodes in a path should be condensed.

Note that our navigational sketch, described in Section 3, maintains all the information that both the above structures maintain: connected components, biconnected components, bridges and articulation points are explicitly managed, and it is easy to see that the nodes in each block component can be derived from the subgraphs obtained by considering only the coloured edges.

Therefore, at a high level view, we can say that the main difference between our approach and the one of [27] is (mainly) whether the biconnectivity information is stored in the colours of the edges or in a distinct type of nodes (round nodes). However, at a lower level, this choice has a big impact on the data structures: the *link/condense* tree of Westbrook and Tarjan, in the two distinct versions for BBF and BF, needs to address properly the two node types, whilst our data structure, detailed in the next section, is built by combining two distinct union-find structures. With respect to (both versions of) the *link/condense tree*, our data structure has the advantage of being simple and easily implementable, whilst capable of achieving (almost) the same bounds as the two approaches of Westbrook and Tarjan: their approach costs $O(m\alpha(m,n))$, whilst ours, as we discuss in Section 7, is slightly worse, i.e., $O(m\alpha(m,n) + n\log n)$ on a sequence of $m$ edges; it is easy to note that, if the graph is slightly more dense than a regular graph ($m/n \geq \log n$), the cost becomes the same.

As already mentioned, we implemented our algorithm and performed an experimental evaluation, detailed in Section 8.

## 5 Proof of Correctness

In order to prove the correctness of *AFL*, we show that it has the following invariants:

1. at each step, each tree in the forest is a connected component (Theorem 1);

2. at each step, each node adjacent to a *c*-colored edge, belongs to the biconnected component represented by *c* (Theorem 2);

3. at each step, each solid edge is a bridge (Theorem 3);

4. at each step, each node $u$ with $d_c(u) > 1$ is an articulation point (Theorem 4);

**Theorem 1** *At each step, each tree in the forest is a connected component.*

**Proof** This holds by induction on the length of the stream of edges: we show now that, after the processing of each edge, the trees in the forest represent the connected components of the graph. The induction base case is before the reading of the stream: the forest has no edges, and each node is a singleton tree, i.e., a singleton connected component; thus the base case is true.

Now let us assume that it holds after the first $k$ edges from the stream, i.e., the trees represent connected components, and let us prove that it holds also after the $(k+1)$-th edge. Indeed, when we see, from the stream, the $(k+1)$-th edge, one of the following situations occurs:

- the $(k+1)$-th edge connects two distinct trees, and these two become a single tree; for the induction hypothesis, the two trees represent two distinct connected components and these two, together with the $(k+1)$-th edge, are merged into a single connected component (of the graph);

- the $(k+1)$-th edge connects two nodes inside the same tree, therefore the only changes can occur inside the tree, no changes in the connected components (of the graph).

It is easy to see that, after both the above cases, the set of trees still represents properly the set of connected components of the graph (seen so far from the stream), and therefore the property still holds. □

**Theorem 2** *At each step, each node adjacent to a c-colored edge, belongs to the biconnected component represented by c.*

**Proof** Also for this case, as we did for the previous theorem, we can provide a proof by induction on the length of the stream of edges. Initially there are no biconnected components (base case), and there are no colored edges; therefore the base case holds.

Let us now assume that it holds after the first $k$ edges from the stream; when we see the $(k+1)$-th edge, one of the following cases occurs:

- the $(k+1)$-th edge connects two distinct trees, and these two become a single tree; no changes in the biconnected components of the graph;

- the $(k+1)$-th edge connects two nodes inside the same tree; either both the nodes have edges of the same color, and therefore no change occurs in the biconnected components (BCCs) of the graph, or nodes connected by solid and distinct colored edges are rearranged into a set of nodes connected by edges of the same color, to match that a new BCC has been found, and this BCC is the merge of several BCCs.

In all the above cases, all the sets of colored edges represent the BCCs of the graph, and therefore the property holds. $\square$

**Theorem 3** *At each step, each solid edge is a bridge.*

**Proof** Also this property can be proved by induction on the length of the stream of edges. As the base case, initially, there are no solid edges, and therefore the base case holds.

As before, let us assume that, after the $k$-th item from the stream all the solid edges are the bridges of the graph. When the $(k+1)$-th item of the stream is processed, one of the following cases happens:

- the $(k+1)$-th edge connects two distinct trees, and these two become a single tree, joined by a solid edge; there is one more bridge now in the graph;

- the $(k+1)$-th edge connects two nodes inside the same tree and both the nodes have edges of the same color: the edge gets dropped and no change occurs in the navigational sketch; no changes in the set of bridges;

- the $(k+1)$-th edge connects two nodes inside the same tree and these nodes have no edges of the same color: the (possibly empty) set of solid edges involved gets replaced by a set of edges of the same color, and a new BCC is found; the (possibly empty) set of bridges inside nodes belonging to this BCC are no longer bridges.

In all the above cases, the property holds. $\square$

**Theorem 4** *At each step, each node $u$ with $d_c(u) > 1$ is an articulation point.*

**Proof** It is easy to see that, from Theorems 2 and 3, the properties of the navigational sketch are maintained, and therefore, due to Lemma 1, $u$ is an articulation point. $\square$

# 6  Data Structure

In this section we analyze the data structure that represents the navigational sketch of a graph $G$. The key issue here is how to deal, given the current edge $(i, j)$ from the stream, with the operations described in Section 4 (please refer to Algorithm 1 and Algorithm 2).

Roughly speaking, the data structure behind the navigational sketch needs to provide information about three distinct aspects:

1. The set of nodes that belong to the same connected component (CC).

2. The set of nodes that belong to the same BCC.

3. The connection between the nodes, represented by paths in the forest.

With the above information, it is easy to see that all the operations performed by the AFL algorithm can be accordingly grouped:

1. Operations related to the Connected Components:

   **op-A.** find whether nodes are in the same tree → $\texttt{find}_{CC}$;

   **op-B.** join trees → $\texttt{union}_{CC}$;

2. Operations related to the Biconnected Components:

   **op-C.** find whether nodes are extremes of same coloured edges → $\texttt{find}_{BCC}$;

   **op-D.** join sets of same coloured edges or solid edges → $\texttt{union}_{BCC}$;

3. An operation related to the paths in the forest:

   **op-E.** find the path joining nodes → $\texttt{LCA}$.

The above information, if disjoint, could be easily implemented with three data structure: two distinct union/find data structures and any suitable data structure able to represent a tree and find paths inside it. The problem arises because we need to combine the three structures into a single one, without affecting the relative performances of each of them. Moreover, the two union/find data structures can not easily be merged because, in order to properly distinguish both connected and biconnected components, one should address the nodes, and the other the edges.

We now present a simple and effective data structure that allows us to perform all the above operations in (almost) optimal time. As shown in Figure 5, for each node of the graph we can store seven ($\log n$ bits) integers. Therefore, we can think of our data structure as a table, whose rows are the nodes and whose columns are, respectively:

1. The father of the node in the forest, if any.

2. The representative element of the biconnected component, if any.

3. The left brother of the node inside the biconnected component, if any.

4. The right brother of the node inside the biconnected component, if any.

5. The representative element of the connected component, if any.

6. The size of the biconnected component.

7. The size of the connected component.

In Table 1 we show an example of such a data structure, that corresponds to the navigational sketch shown in the left side of Figure 5; we also show a graphical (pointer) view of the contents of the first four columns of the table, in order to allow an easy comparison between the Navigational Sketch and its array representation. In particular, we can note the following properties of this structure:

- we use a *rooted tree* to represent path and for each node, its father is stored in the *father* column if it is the representative of its BCC.

- we use a *tree* representation for the CC disjoint-set and for each node, its CC representative (the root of the tree) is stored in the *CC-representative* column if it is the representative of its BCC.

- we use a *linked-list* representation for the BCC disjoint-set and all edge information is stored in the child node row, in particular in the *left-brother*, *right-brother* and *BCC-representative* columns.

- to implement the union-by-size disjoint set, we store the size of CCs and BCCs respectively in their representative rows.

Note that, in the table, only the relevant values are shown, while the others are represented with a dash ($-$); in the actual execution of the algorithm we allow some spurious data to remain in the table. For example, the father of a node $v$ is the father of the BCC representative of $v$ (i.e., `father[BCCrep[v]]`), and therefore `father[v]` can assume any value in the table: e.g., in Table 1, the father of node 4 is the father of its BCC representative (node 3) and, therefore, it is node 2; this way, for example, when we merge two BCCs we do not need to update the father value of each node.

| Node | Father | BCC Rep | Left Brother | Right Brother | CC Rep. | BCC Size | CC Size |
|------|--------|---------|--------------|---------------|---------|----------|---------|
| 1 | -1 | 1 | 1 | 1 | 1 | 1 | 6 |
| 2 | 1  | 2 | 2 | 2 | 1 | 1 | - |
| 3 | 2  | 3 | 3 | 4 | 1 | 3 | - |
| 4 | -  | 3 | 3 | 4 | - | - | - |
| 5 | -1 | 5 | 5 | 5 | 5 | 1 | 1 |
| 6 | 4  | 6 | 6 | 7 | 1 | 3 | - |
| 7 | -  | 6 | 6 | 7 | - | - | - |

Table 1: Array data relative to the navigational sketch shown in Figure 5 (left); a graphical (pointer) view of the first four columns of this table is depicted in the right side of Figure 5.

We conclude this section with a brief discussion about the space complexity of this algorithm. As we noticed, we keep in main memory only the navigational sketch, and therefore the memory occupation is $O(n \log n)$ bits. It is important to note that, if the input graph is a tree, then all its edges are bridges, and therefore the space needed to store part of the solution, i.e., the list of all the bridges, is $\Omega(n \log n)$; thus the space occupation of AFL is tight.

## 7 Analysis

Let us begin the analysis by considering the cost of processing the whole stream, in terms of the maximum number of times that each *macro* operation, i.e., Least Common Ancestor (`LCA`), `find` and `union`, can be executed. If the graph has $m$ edges and $n$ nodes, we have:

1. $2m$ $\text{find}_{CC}$, one for each edge of the stream;

2. $2m$ $\text{find}_{BCC}$, one for each edge for the stream (minus some edges needed to build initially some biconnected components, but let us not count

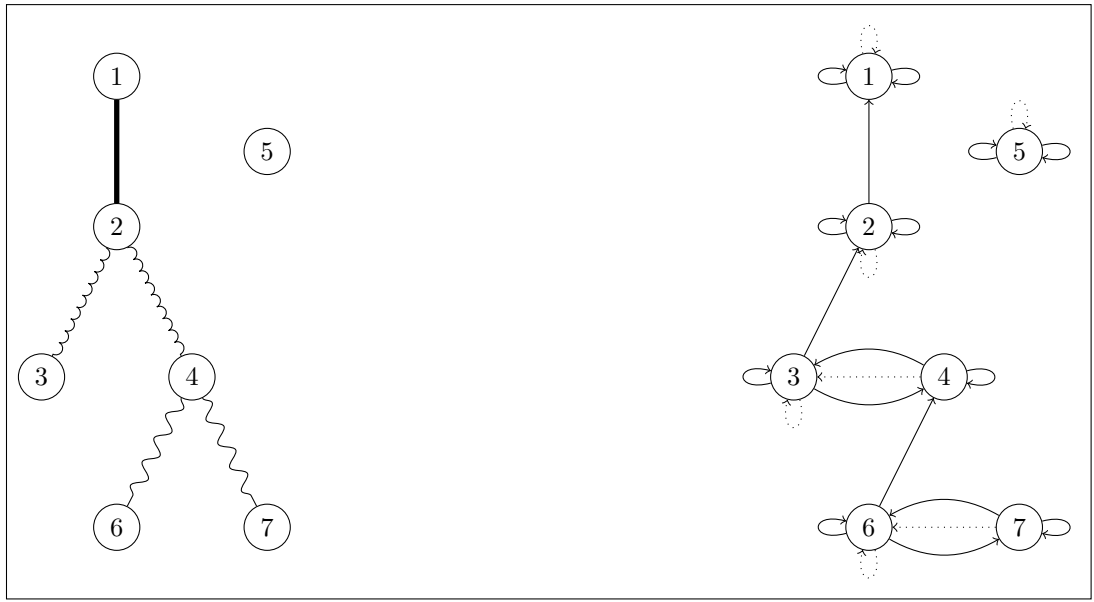

Figure 5: Example of a Navigational Sketch (left side) and a graphical (pointer) view of the (first four columns of the) array data from Table 1 (right side): here, for each node, dotted edges point to the BCC representative, vertical edges points to the father, and lateral edges point to the left and right brother. Note that, for the navigational sketch, as for Figure 2, with the line style ▬▬ we represent solid edges, while colored ones are represented by the styles ⌒⌒⌒ , and ⌒⌒ .

them);

3. $n-1$ $\text{union}_{CC}$, because a node, once inserted in a connected component, cannot be removed;

4. $n-1$ $\text{union}_{BCC}$, because a node, once inserted in a biconnected component, cannot be removed;

5. $n-1$ LCA, to identify which BCCs are to be unioned;

Summing up, we have, for the whole sequence:

$$2m(\text{find}_{CC} + \text{find}_{BCC}) + \cdots$$

$$\cdots + (n-1)(\text{union}_{CC} + \text{union}_{BCC} + \text{LCA})$$

We now introduce in the analysis the costs of LCA, find and union operations. The upper bounds are the following:

**CC.** implementing union-by-size and path-compression heuristics, it is known [7] that the cost, for a sequence of $m$ finds on $n$ elements and $n-1$ unions, is $O(n + m\alpha(m, n))$, where $\alpha$ is a very slowing growing function, the inverse Ackermann function.

**BCC.** we used a union-by-size heuristic on a linked list with path compression that, as above, on a sequence of $m$ finds on $n$ elements and $n-1$ unions, has a cost of $O(n + m\alpha(m, n))$ [7].

**LCA.** in order to find the least common ancestor between two nodes, we go up from both of them, marking every visited node until we find an already marked node; the cost of this operation is therefore $O(d)$ where $d$ is the max depth of the two nodes relative to their LCA. Note that, all the nodes in the path are moved to the same level of the tree, and therefore its amortized cost is constant and, on a sequence of $n-1$ LCAs, is $O(n)$.

It should be clear, from the data structure presented in the previous section, that the BCC and CC union/find structure are not the same: in particular, when we merge two distinct CC, we need to evert the smaller tree[2]. This adds a cost of $O(n \log n)$ for a sequence of $n-1$ unions.

Summing up, it easily follows:

---

[2]To be more precise, if the smaller tree has $k$ elements and a depth of $d$, we do not need to update the whole tree ($k$ elements), but only $d$ elements, one for each level of the tree.

**Theorem 5** *The worst case overall processing time is $O(m\alpha(m,n) + n\log n)$.*

Finally, if we divide the cost of processing the whole stream by the number of its elements, that is $m$, we have the amortized per item processing time.

**Theorem 6** *The amortized per item processing time is $O(\frac{n}{m}\log n + \alpha(m,n)))$.*

**Corollary 1** *If the average degree of the graph is greater than or equal to $\log n$, then the amortized per item processing time is $O(\alpha(m,n))$.*

In other words, the above corollary shows that, if the graph is dense enough ($m/n \geq \log n$), the amortized PIPT is almost constant. We will see in the following section that this result is confirmed in our experimental analysis.

## 8  Experimental Results

In this section we present the results of a brief experimental validation of the AFL algorithm: we focus on a few aspects aimed at showing that our approach is sound and performs effectively, with an amortized time that, in practice, is independent of the number of edges $m$.

All the experiments described were performed on an off-the-shelf computer: a dual boot (Windows Vista and Ubuntu Linux) laptop Dell XPS M1330 (4Gb RAM, Intel Core2 Duo T8100 2.1GHz). We implemented AFL in the C programming language, using the `gcc` compiler under both Windows and Ubuntu. We did not observe significant differences between the two operating systems, and the times reported are an average of 10 runs for each operating system. Our datasets, as shown in Table 4 where we report also the relative data repositories, are real-world graphs freely available, including Autonomous System graphs, Web graphs, and graphs from several application domains; the (simple) source code of the AFL algorithm is available at the address `www.dis.uniroma1.it/~firmani/afl`.

The Autonomous System graphs are the original motivation for our work: we wanted to develop a (real-time) streaming algorithm, able to compute all the (bi)connectivity properties on this graph (that is the backbone of the Internet), listening to BGP announcements (the stream). Even if AS graphs are very sparse, the performances of AFL algorithm are more than encouraging and very similar between all the several instances we tested; therefore, in Table 2 we report only the results on one sample of this network.

| Graph | Type | Data Source | Disk Space | Number of Nodes: $n = |V|$ | Number of Edges: $m = |E|$ | Average Degree: $\frac{m}{n}$ | Density Factor: $\frac{m}{n \log n}$ | Max # *touched* edges per operation | Avg. # *touched* edges per operation | Overall Processing Time (t) | Amortized PIPT: $\frac{t}{m}$ | Edges processed per second: $\frac{m}{t}$ |
|---|---|---|---|---|---|---|---|---|---|---|---|---|
| Yeast | biology | 3 | 70 Kb | 7.1k | 6.6k | 0.93 | 13.82 | 6 | 0.55 | < 0.01 | - | - |
| DutchElite | economy | 3 | 54 Kb | 4.7k | 5.2k | 1.10 | 11.11 | 8 | 1.04 | < 0.01 | - | - |
| email | social | 1 | 93 Kb | 1.1k | 10.7k | 9.43 | 1.08 | 2 | 0.20 | < 0.01 | - | - |
| USpowerGrid | technology | 3 | 136 Kb | 4.9k | 9.9k | 2.00 | 6.13 | 15 | 0.79 | < 0.01 | - | - |
| AS | Aut. Systems | 4 | 670 Kb | 65.5k | 57.4k | 0.88 | 18.27 | 4 | 0.78 | < 0.1 | ≈ 0.52E-6 | - |
| PairsP | similarity | 3 | 737 Kb | 10.6k | 72.0k | 6.78 | 1.97 | 6 | 0.24 | < 0.1 | ≈ 0.43E-6 | - |
| dic-28 | linguistic | 3 | 1 Mb | 52.6k | 89k | 1.69 | 9.28 | 9 | 0.64 | < 0.1 | ≈ 0.56E-6 | - |
| foldoc | linguistic | 3 | 1.3 Mb | 13.3k | 119.8k | 8.97 | 1.53 | 6 | 0.11 | < 0.1 | ≈ 0.67E-6 | - |
| wordnet3 | linguistic | 3 | 1.6 Mb | 82.6k | 124.7k | 1.51 | 10.83 | 8 | 0.87 | < 0.1 | ≈ 0.67E-6 | - |
| eatRS | linguistic | 3 | 3.8 Mb | 23.2k | 325.0k | 14.00 | 1.05 | 9 | 0.13 | < 0.2 | ≈ 0.38E-6 | - |
| hep-th-new | citation | 2 | 4.2 Mb | 27.7k | 352.7k | 12.70 | 1.16 | 7 | 0.15 | < 0.2 | ≈ 0.50E-6 | - |
| cnr-2000 | web | 5 | 44.7 Mb | 325k | 3.2M | 9.88 | 1.85 | 9 | 0.18 | < 3 | ≈ 0.80E-6 | ≈ 1M |
| eu-2005 | web | 5 | 270.8 Mb | 862k | 19.2M | 22.3 | 0.88 | 7 | 0.08 | < 10 | ≈ 0.51E-6 | ≈ 2.0M |
| indochina-2004 | web | 5 | 3 Gb | 7.4M | 194.1M | 26.18 | 0.87 | 72 | 0.07 | < 100 | ≈ 0.47E-6 | ≈ 2.1M |
| uk-2002 | web | 5 | 5 Gb | 18.5M | 290.6M | 15.70 | 1.54 | 91 | 0.12 | < 160 | ≈ 0.54E-6 | ≈ 1.8M |
| it-2004 | web | 5 | 20.5 Gb | 41.2M | 1.1G | 27.42 | 0.92 | 225 | 0.07 | < 610 | ≈ 0.54E-6 | ≈ 1.8M |

Table 2: Experimental results; time expressed in seconds. The last column presents the values only for the graphs whose overall processing time was greater than one second.

Later, we wanted to test the scalability of our approach, and therefore we analyzed several samples of *Web* graphs, that are, to the best of our knowledge, the biggest graphs freely available in the net: they are samples of the link structure of Web pages. A Web graph is a graph whose nodes are the static pages of the Web, and whose (directed) edges are the links between them. The main ingredient of the Google ranking system is the PageRank algorithm [21], that uses the graph structure of the pages, i.e., the Web graph, to derive their relative ranking; after the success of PageRank, the properties of the Web graph have been studied extensively, starting from the seminal work of Broder et al. [5]. Note that, when dealing with Web graphs, we considered their links as undirected.

Finally, in order to provide a more complete analysis, we also tested the AFL algorithm against graphs from many application domains, and with different density features; note, however, that none of these graphs, can be compared, in size, with the huge samples of the Web graphs. The details about these additional graphs can be found in Table 3, whilst Table 4 provides a list of their sources.

The results are shown in Table 2; as we already mentioned in the worst case analysis (see Corollary 1), algorithm AFL performs better if the graph is dense enough ($m/n \geq \log n$ and therefore when $\frac{n \log n}{m} \leq 1$). The theoretical analysis ensures that, for a particular network, the more this value that we call *density factor* is smaller than 1, the better the algorithm works. Therefore analysis against sparse networks seemed to us more interesting.

In Table 2 the graphs are ordered by their size (and, therefore, by the number of edges) and, for each graph, we first report its type together with its source. Then, we provide few statistics: in particular, we can see the disk space occupation (in order to normalize the graphs from distinct sources, stored with distinct representations, we report the size of the text file containing the list of edges, where each node is identified with an increasing number in the range $0..n-1$), the number of nodes, the number of edges, the average degree, and a density factor $\frac{n \log n}{m}$. In Appendix B, for each graph we detail these results by providing the same statistics on four distinct streams made, respectively, by 25%, 50%, 75%, and 100% of the overall stream (Tables 5-8).

Finally, the last five columns of Table 2 report a few parameters about the AFL execution: in particular, we see the overall processing time, and the amortized PIPT, i.e., the overall processing time divided by the number of edges (the length of the stream). We also show the inverse of the amortized

PIPT, that is the number of edges processed per second (in the table, it is shown only for the graphs whose overall processing time was greater than one second). We also measured, for each operation, the number of edges in the navigational sketch that were modified. We report, for each graph, both the maximum and the average value of this measure: the first one representing an evaluation of the worst case PIPT in the same execution, in order to test how the AFL algorithm could deal with a real-time execution, and the second one representing the average PIPT in terms of modified edges.

Looking at Table 2, it is interesting to point out that, on an average laptop, the algorithm is able to process more than one million edges per second (see the last column in Table 2): this seems more than reasonable even with a real-time constraint. Furthermore, as already mentioned, the values of the ratio $t/m$ seem to suggest that the per item processing time is, in practice, almost constant: for big enough graphs the order of magnitude increases at most by one (and not as a function of $m$), with $m$ ranging from 10k to 1.1G.

The different performances on the graphs seem to depend on their structural properties and, in particular, on the structure of the navigational sketch: intuition suggests that, if the graph contains many BCCs, these tend to "collapse" into subtrees of unitary height (i.e., one father with its sons), and after the initial construction of the subtree all the edges inside the BCCs are "dropped" quickly.

We conclude by observing that, according to the measured values, differently from the theoretical results, in practice there is no correlation between the density factor and the amortized PIPT. Also, there is no evident correlation between the number of edges of the graph and the maximum number of edges modified in the navigational sketch in a single operation; on the contrary, we observe that the average number of modified edges in the navigational sketch seems to be slightly decreasing as the number of edges increases; however, to further investigate this phenomenon, a more detailed experimental evaluation is needed.

## 9 Conclusions

In this paper we presented *At First Look*, an algorithm to real-time monitor an undirected network by computing all its (bi)connectivity properties: articulation points, bridges, connected and biconnected components.

In order to implement the AFL algorithm, we designed a data structure, simple to implement and effective. This data structure represents what we called *Navigational Sketch*, and provides all the (bi)connectivity properties of the input graph; furthermore, this Navigational Sketch can be used as a building block in the development of more complex streaming graph algorithms.

In the paper we provide a correctness proof and complexity analysis of the algorithm showing that its memory space occupation, $O(n \log n)$, is tight (Section 6). We also proved a bound on the per item processing time, and presented the results of an experimental evaluation of our algorithm against real-world networks, with up to 1.1G edges.

This experimental evaluation confirms the effectiveness of our approach, and on an off-the-shelf laptop (with CPU able to perform two billion clock cycles per second) we are able to process more than one million edges per second; this effectiveness is confirmed also from a different point of view: the number of edges modified in the navigational sketch, on the average, is less than one per each edge read from the stream.

# A   An example of AFL execution

In this section we see an example of the execution of At First Look (AFL). The input graph is shown in Figure 6; it has 11 nodes and 15 edges. The label on each edge represents the order of the edge inside the stream.

Figure 7 shows the AFL execution: here, we see the snapshots of the forest after the processing of each edge in the stream. Note that, in order to allow black and white printing we represent solid and colored edges with different line styles.

The final configuration of the forest is shown in the bottommost, rightmost tree (after the processing of the last edge): here we see that the graph is connected, i.e., it contains a single connected component, has two bridges ($(E, G)$ and $(F, K)$), three biconnected components (respectively, $(A, B, C, D, E)$, $(G, H, F)$, and $(H, I, J)$). The articulation points are:

- $E$, $F$, and $G$, because they are all adjacent to a bridge and a biconnected component.

- $H$, because it belongs to two distinct biconnected components.

Note that $K$, which is adjacent to a bridge, is not an articulation point because its removal does not disconnect the graph.

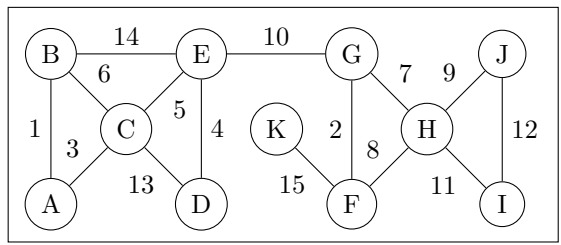

Figure 6: Input graph.

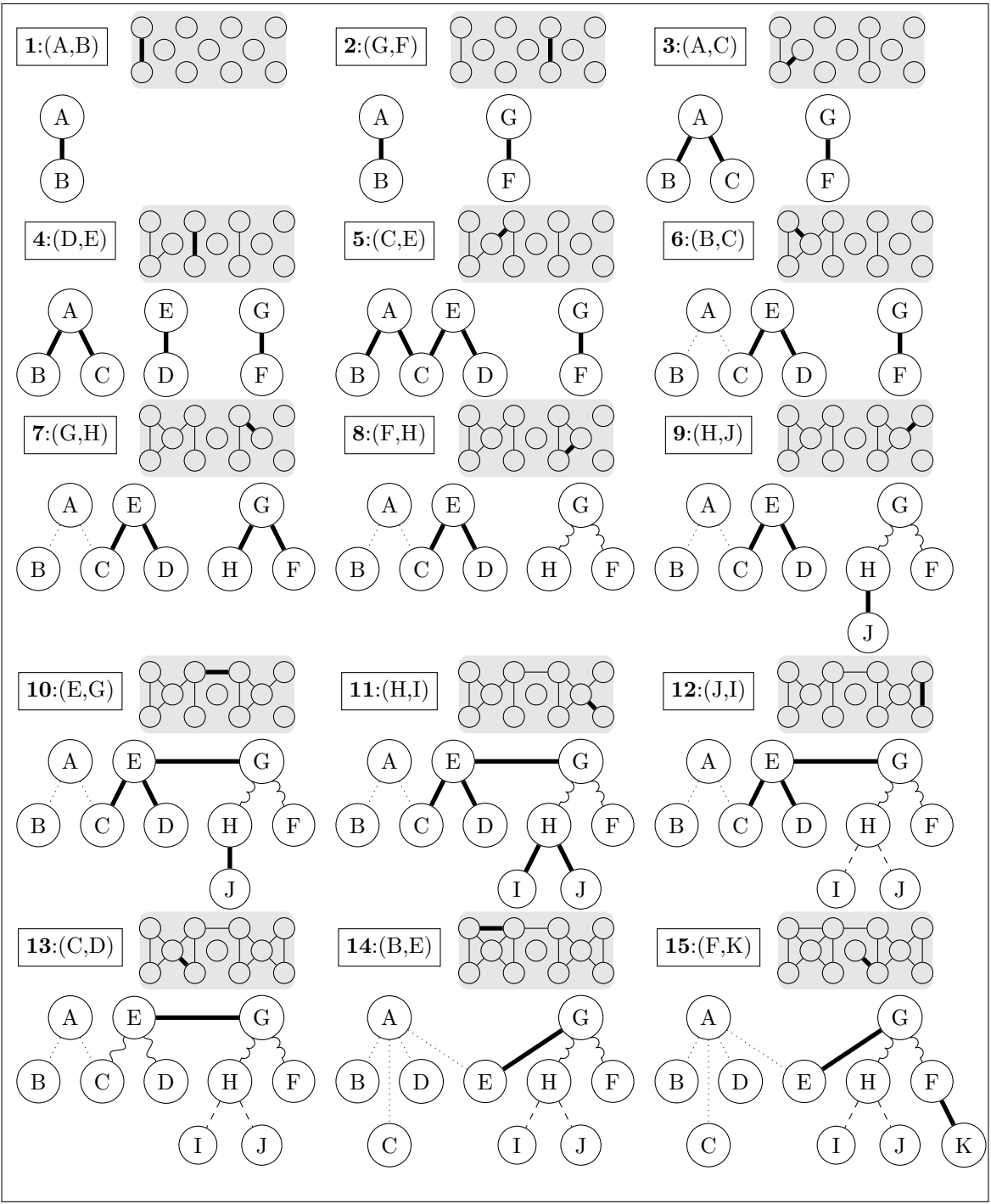

Figure 7: Snapshots of AFL execution against the input graph shown in Figure 6; the current edge from the stream is depicted in bold inside the small input graph; each snapshot shows the state of the forest after each edge: with the line style ——— we represent solid edges, while colored ones are represented, respectively, by the styles ········, ∿∿, ⌇⌇, and - - - -.

# B  Experimental results details: dataset descriptions, repositories and partial stream computations

In this section we provide some details about the experimental results; in particular, in Table 3 we provide a brief description of each graph, whilst in Table 4 we list their sources. We conclude by detailing, from Table 5 to Table 8, the results obtained, for each graph, when 25%, 50%, 75%, and 100% of the graph is processed.

| Graph | Repository # | Description |
|---|---|---|
| Yeast | 3 | Protein-protein interaction network described and analyzed in budding yeast |
| DutchElite | 3 | Data on the administrative elite in The Netherlands |
| email | 1 | Network of e-mail interchanges between members of the Univeristy Rovira i Virgili (Tarragona) |
| AS | 4 | Network of the Autonomous Systems |
| PairsP | 3 | Similar to eatRS but with associations collected during a different experiment |
| dic-28 | 3 | English words from Knuth's dictionary with 2 to 8 characters. An edge (X,Y) from term X to term Y exists iff the term Y can be reached changing, removing or adding a single character from the term X |
| foldoc | 3 | Free On-line Dictionary of Computing (FOLDOC) is an online encyclopedic dictionary of computing subjects. An edge (X,Y) from term X to term Y exists if the term Y is used to describe the meaning of term X |
| wordnet3 | 3 | WordNet is a lexical database for the English language. Edge set is built as in foldoc |
| eatRS | 3 | The Edinburgh Associative Thesaurus (EAT) is a set of word association norms showing the counts of word association as collected from subjects |
| hep-th-new | 2 | Network of coauthorships between scientists posting preprints on the High-Energy Theory E-Print Archive between Jan 1, 1995 and December 31, 1999 |
| cnr-2000 | 5 | Crawl of the Italian CNR domain in 2000 |
| eu-2005 | 5 | Crawl of the .eu domain performed in 2005 |
| indochina-2004 | 5 | Crawl collected in 2004 with the support of the Language Observatory Project |
| uk-2002 | 5 | Crawl of the .uk domain performed by UbiCrawler in 2002 |
| it-2004 | 5 | Crawl of the .it domain performed in 2004 |

Table 3: Dataset description. The repository number refers to Table 4.

| # | Author | URL |
|---|---|---|
| 1 | A. Arenas | `http://deim.urv.cat/~aarenas/data/welcome.htm` |
| 2 | M. Newman | `http://www-personal.umich.edu/~mejn/netdata/` |
| 3 | Pajek datasets (V. Batagelj and A. Mrvar) | `http://vlado.fmf.uni-lj.si/pub/networks/data/` |
| 4 | Route Views | `http://www.routeviews.org/` |
| 5 | P. Boldi and S. Vigna [4] | `http://law.dsi.unimi.it/` |

Table 4: Dataset repositories.

| Graph | Type | % | Disk Space | Number of Nodes: $n = \|V\|$ | Number of Edges: $m = \|E\|$ | Average Degree: $\frac{m}{n}$ | Density Factor: $\frac{m}{n \log n}$ | Max # *touched* edges per operation | Avg. # *touched* edges per operation | Overall Processing Time (t) | Amortized PIPT: $\frac{t}{m}$ | Edges processed per second: $\frac{m}{t}$ |
|---|---|---|---|---|---|---|---|---|---|---|---|---|
| Yeast | biology | 25% | 70 Kb | 7.1k | 1.7k | 0.24 | 53.88 | 6 | 0.75 | < 0.01 | - | - |
| | | 50% | | | 3.4k | 0.47 | 27.16 | 6 | 0.62 | < 0.01 | - | - |
| | | 75% | | | 5.0k | 0.70 | 18.33 | 6 | 0.57 | < 0.01 | - | - |
| | | 100% | | | 6.6k | 0.93 | 13.82 | 6 | 0.55 | < 0.01 | - | - |
| DutchElite | economy | 25% | 54 Kb | 4.7k | 1.3k | 0.27 | 44.49 | 8 | 1.11 | < 0.01 | - | - |
| | | 50% | | | 2.6k | 0.55 | 22.23 | 8 | 1.07 | < 0.01 | - | - |
| | | 70% | | | 4.0k | 0.82 | 14.81 | 8 | 1.06 | < 0.01 | - | - |
| | | 100% | | | 5.2k | 1.10 | 11.11 | 8 | 1.04 | < 0.01 | - | - |
| email | social | 25% | 93 Kb | 1.1k | 2.7k | 2.37 | 4.28 | 2 | 0.45 | < 0.01 | - | - |
| | | 50% | | | 5.4k | 4.76 | 2.13 | 2 | 0.30 | < 0.01 | - | - |
| | | 70% | | | 8.1k | 7.15 | 1.42 | 2 | 0.24 | < 0.01 | - | - |
| | | 100% | | | 10.7k | 9.43 | 1.08 | 2 | 0.20 | < 0.01 | - | - |
| USpowerGrid | technology | 25% | 136 Kb | 4.9k | 2.5k | 0.50 | 24.43 | 11 | 0.87 | < 0.01 | - | - |
| | | 50% | | | 5.0k | 1.01 | 12.11 | 11 | 0.83 | < 0.01 | - | - |
| | | 70% | | | 7.5k | 1.51 | 8.12 | 15 | 0.80 | < 0.01 | - | - |
| | | 100% | | | 9.9k | 2.00 | 6.13 | 15 | 0.79 | < 0.01 | - | - |

Table 5: Experimental results, part 1 of 4; time expressed in seconds. The last column presents the values only for the graphs whose overall processing time was greater than one second.

| Graph | Type | % | Disk Space | Number of Nodes: $n = |V|$ | Number of Edges: $m = |E|$ | Average Degree: $\frac{m}{n}$ | Density Factor: $\frac{m}{n \log n}$ | Max # *touched* edges per operation | Avg. # *touched* edges per operation | Overall Processing Time (t) | Amortized PIPT: $\frac{t}{m}$ | Edges processed per second: $\frac{m}{t}$ |
|---|---|---|---|---|---|---|---|---|---|---|---|---|
| AS | Aut. Systems | 25% | 670 Kb | 65.5k | 14.3k | 0.22 | 73.24 | 4 | 0.67 | < 0.1 | ≈ 0.84E-6 | - |
| | | 50% | | | 28.6k | 0.44 | 36.60 | 4 | 0.72 | < 0.1 | ≈ 0.56E-6 | - |
| | | 75% | | | 43.0k | 0.66 | 24.38 | 4 | 0.75 | < 0.1 | ≈ 0.53E-6 | - |
| | | 100% | | | 57.4k | 0.88 | 18.27 | 4 | 0.78 | < 0.1 | ≈ 0.52E-6 | - |
| PairsP | similarity | 25% | 737 Kb | 10.6k | 18.0k | 1.70 | 7.90 | 6 | 0.40 | < 0.1 | ≈ 0.78E-6 | - |
| | | 50% | | | 36.0k | 3.39 | 3.95 | 6 | 0.25 | < 0.1 | ≈ 0.42E-6 | - |
| | | 75% | | | 54.0k | 5.08 | 2.63 | 6 | 0.18 | < 0.1 | ≈ 0.44E-6 | - |
| | | 100% | | | 72.0k | 6.78 | 1.97 | 6 | 0.24 | < 0.1 | ≈ 0.43E-6 | - |
| dic-28 | linguistic | 25% | 1 Mb | 52.6k | 22.3k | 0.42 | 37.10 | 7 | 0.29 | < 0.1 | ≈ 0.94E-6 | - |
| | | 50% | | | 44.5k | 0.84 | 18.55 | 7 | 0.60 | < 0.1 | ≈ 0.58E-6 | - |
| | | 75% | | | 66.8k | 1.27 | 12.37 | 7 | 0.64 | < 0.1 | ≈ 0.57E-6 | - |
| | | 100% | | | 89k | 1.69 | 9.28 | 9 | 0.64 | < 0.1 | ≈ 0.56E-6 | - |
| foldoc | linguistic | 25% | 1.3 Mb | 13.3k | 29.7k | 2.22 | 6.17 | 6 | 0.40 | < 0.1 | ≈ 0.58E-6 | - |
| | | 50% | | | 59.7k | 4.47 | 3.06 | 6 | 0.30 | < 0.1 | ≈ 0.32E-6 | - |
| | | 75% | | | 89.8k | 6.72 | 2.04 | 6 | 0.26 | < 0.1 | ≈ 0.36E-6 | - |
| | | 100% | | | 119.8k | 8.97 | 1.53 | 6 | 0.11 | < 0.1 | ≈ 0.67E-6 | - |

Table 6: Experimental results, part 2 of 4; time expressed in seconds. The last column presents the values only for the graphs whose overall processing time was greater than one second.

| Graph | Type | % | Disk Space | Number of Nodes: $n = \|V\|$ | Number of Edges: $m = \|E\|$ | Average Degree: $\frac{m}{n}$ | Density Factor: $\frac{m}{n \log n}$ | Max # *touched* edges per operation | Avg. # *touched* edges per operation | Overall Processing Time ($t$) | Amortized PIPT: $\frac{t}{m}$ | Edges processed per second: $\frac{m}{t}$ |
|---|---|---|---|---|---|---|---|---|---|---|---|---|
| wordnet3 | linguistic | 25% | 1.6 Mb | 82.6k | 33.3k | 0.40 | 40.51 | 7 | 0.93 | < 0.1 | ≈ 0.81E-6 | - |
| | | 50% | | | 66.7k | 0.81 | 20.25 | 7 | 0.94 | < 0.1 | ≈ 0.72E-6 | - |
| | | 75% | | | 97.4k | 1.18 | 13.87 | 8 | 0.92 | < 0.1 | ≈ 0.55E-6 | - |
| | | 100% | | | 124.7k | 1.51 | 10.83 | 8 | 0.87 | < 0.1 | ≈ 0.67E-6 | - |
| eatRS | linguistic | 25% | 3.8 Mb | 23.2k | 81.3k | 3.50 | 4.14 | 9 | 0.29 | < 0.1 | ≈ 0.60E-6 | - |
| | | 50% | | | 162.5k | 7 | 2.07 | 9 | 0.18 | < 0.1 | ≈ 0.52E-6 | - |
| | | 75% | | | 243.8k | 10.50 | 1.38 | 9 | 0.15 | < 0.2 | ≈ 0.42E-6 | - |
| | | 100% | | | 325.0k | 14.00 | 1.05 | 9 | 0.13 | < 0.2 | ≈ 0.38E-6 | - |
| hep-th-new | citation | 25% | 4.2 Mb | 27.7k | 88.1k | 3.17 | 4.65 | 7 | 0.28 | < 0.1 | ≈ 0.50E-6 | - |
| | | 50% | | | 176.3k | 6.35 | 2.32 | 7 | 0.21 | < 0.1 | ≈ 0.38E-6 | - |
| | | 75% | | | 264.5k | 9.52 | 1.55 | 7 | 0.18 | < 0.2 | ≈ 0.46E-6 | - |
| | | 100% | | | 352.7k | 12.70 | 1.16 | 7 | 0.15 | < 0.2 | ≈ 0.50E-6 | - |
| cnr-2000 | web | 25% | 44.7 Mb | 325k | 0.8M | 2.33 | 7.85 | 9 | 0.30 | < 0.5 | ≈ 0.54E-6 | ≈ 1.8M |
| | | 50% | | | 1.5M | 4.71 | 3.89 | 9 | 0.23 | < 0.9 | ≈ 0.53E-6 | ≈ 2.0M |
| | | 75% | | | 2.3M | 7.08 | 2.59 | 9 | 0.20 | < 2 | ≈ 0.48E-6 | ≈ 2.1M |
| | | 100% | | | 3.2M | 9.88 | 1.85 | 9 | 0.18 | < 3 | ≈ 0.80E-6 | ≈ 1M |

Table 7: Experimental results, part 3 of 4; time expressed in seconds. The last column presents the values only for the graphs whose overall processing time was greater than one second.

| Graph | Type | % | Disk Space | Number of Nodes: $n = \|V\|$ | Number of Edges: $m = \|E\|$ | Average Degree: $\frac{m}{n}$ | Density Factor: $\frac{m}{n \log n}$ | Max # *touched* edges per operation | Avg. # *touched* edges per operation | Overall Processing Time (t) | Amortized PIPT: $\frac{t}{m}$ | Edges processed per second: $\frac{m}{t}$ |
|---|---|---|---|---|---|---|---|---|---|---|---|---|
| eu-2005 | web | 25% | 270.8 Mb | 862k | 4.7M | 5.41 | 3.64 | 5 | 0.1 | < 3 | ≈ 0.47E-6 | ≈ 2.1M |
|  |  | 50% |  |  | 9.3M | 10.83 | 1.82 | 6 | 0.09 | < 5 | ≈ 0.46E-6 | ≈ 2.2M |
|  |  | 75% |  |  | 14.1M | 16.33 | 1.21 | 7 | 0.08 | < 7 | ≈ 0.47E-6 | ≈ 2.1M |
|  |  | 100% |  |  | 19.2M | 22.3 | 0.88 | 7 | 0.08 | < 10 | ≈ 0.51E-6 | ≈ 2.0M |
| indochina-2004 | web | 25% | 3 Gb | 7.4M | 4.8M | 6.50 | 3.51 | 21 | 0.03 | < 22 | ≈ 0.44E-6 | ≈ 2.3M |
|  |  | 50% |  |  | 95.1M | 12.93 | 1.76 | 57 | 0.05 | < 42 | ≈ 0.44E-6 | ≈ 2.3M |
|  |  | 75% |  |  | 143.2M | 19.32 | 1.18 | 58 | 0.06 | < 57 | ≈ 0.40E-6 | ≈ 2.5M |
|  |  | 100% |  |  | 194.1M | 26.18 | 0.87 | 72 | 0.07 | < 100 | ≈ 0.47E-6 | ≈ 2.1M |
| uk-2002 | web | 25% | 5 Gb | 18.5M | 72.6M | 3.92 | 6.16 | 24 | 0.12 | < 30 | ≈ 0.40E-6 | ≈ 2.5M |
|  |  | 50% |  |  | 145.2M | 7.84 | 3.08 | 91 | 0.12 | < 59 | ≈ 0.40E-6 | ≈ 2.5M |
|  |  | 75% |  |  | 217.8M | 11.76 | 2.05 | 91 | 0.12 | < 86 | ≈ 0.39E-6 | ≈ 2.5M |
|  |  | 100% |  |  | 290.6M | 15.70 | 1.54 | 91 | 0.12 | < 160 | ≈ 0.54E-6 | ≈ 1.8M |
| it-2004 | web | 25% | 20.5 Gb | 41.2M | 282.3M | 6.84 | 3.70 | 225 | 0.08 | < 115 | ≈ 0.41E-6 | ≈ 2.5M |
|  |  | 50% |  |  | 565.6M | 13.70 | 1.85 | 225 | 0.07 | < 217 | ≈ 0.38E-6 | ≈ 2.6M |
|  |  | 75% |  |  | 848.5M | 20.55 | 1.23 | 225 | 0.07 | < 353 | ≈ 0.42E-6 | ≈ 2.4M |
|  |  | 100% |  |  | 1.1G | 27.42 | 0.92 | 225 | 0.07 | < 610 | ≈ 0.54E-6 | ≈ 1.8M |

Table 8: Experimental results, part 4 of 4; time expressed in seconds. The last column presents the values only for the graphs whose overall processing time was greater than one second.